\documentclass[12pt]{article}

\def\be{\begin{equation}}
\def\ee{\end{equation}}

\usepackage{euscript,latexsym}

\begin{document}

\title{Bell's Theorem and Locality in Space}
\author{Igor V. Volovich\thanks{Permanent address: Steklov Mathematical
Institute, Gubkin St.8, GSP-1, 117966, Moscow, Russia;
volovich@mi.ras.ru}\\~\\ {\it Department of Mathematics,
Statistics and Computer Sciences}\\ {\it University of Vaxjo,
S-35195, Sweden}\\  {\it igor.volovich@msi.vxu.se} }
\date{}
\maketitle

\begin{abstract}

Bell's theorem states that some quantum correlations can not be
represented by classical correlations of separated random
variables. It has been interpreted as incompatibility of the
requirement of locality with quantum mechanics. We point out that
in fact the space part of the wave function was neglected in the
proof of Bell's theorem. However this space part is crucial for
considerations of property of locality of quantum system. Actually
the space part leads to an extra factor in quantum correlations
and as a result the ordinary proof of Bell's theorem fails in this
case. We present a criterium of locality in a realist theory of
hidden variables. It is argued that predictions of quantum
mechanics for Gaussian wave functions can be consistent with
Bell's inequalities  and hence Einstein's local realism is
restored in this case.
\end{abstract}

\newpage
\section{Introduction}

Bell's theorem \cite{Bel} states that there are quantum
correlation functions that can not be represented as classical
correlation functions of separated  random variables. It has been
interpreted as incompatibility of the requirement of locality with
the statistical predictions of quantum mechanics \cite{Bel}. For a
recent discussion of Bell's theorem see, for example
 \cite{Hom} - \cite{Gis} and references therein.
 It is now widely accepted, as a result of
Bell's theorem and related experiments, that "local realism" must
be rejected.

Evidently,  the very formulation of the problem of locality in
quantum mechanics is based on ascribing a special role to the
position in ordinary three-dimensional space. It is rather
surprising therefore that the space dependence of the wave
function is neglected in   discussions of the problem of locality
in relation to Bell's inequalities. Actually it is the space part
of the wave function which is relevant to the consideration of the
problem of locality.

In this note we point out that the space part of the wave function
leads to an extra factor in quantum correlation and as a result
the ordinary proof of Bell's theorem fails in this case. We
present a criterium of locality (or nonlocality) of quantum theory
in a realist model of hidden variables. We argue that predictions
of quantum mechanics can be consistent with   Bell's inequalities
for Gaussian wave functions and hence Einstein's local realism is
restored in this case.

\section {Bell's Theorem}

Consider a pair of spin one-half particles formed in the singlet
spin state and moving freely in opposite directions. If one
neglects the space part of the wave function  then the quantum
mechanical correlation of two spins in the singlet state
$\psi_{spin}$ is
\be
\label{eqn1}
 E_{spin}(a,b)=\left<\psi_{spin}|\sigma\cdot a \otimes\sigma\cdot
b|\psi_{spin}\right>=-a\cdot b
 \ee
 Here $a$ and $b$ are two
unit vectors in three-dimensional space and
$\sigma=(\sigma_1,\sigma_2,\sigma_3)$ are the Pauli matrices.
Bell's theorem states that the function $ E_{spin}(a,b)$
(\ref{eqn1})
 can not be represented in the
form
\be
\label{eqn2} P(a,b)=\int A(a,\lambda) B(b,\lambda) d\rho(\lambda)
\ee
 Here $ A(a,\lambda)$ and $  B(b,\lambda)$ are random fields on the sphere,
 $| A(a,\lambda)|\leq 1$, $ | B(b,\lambda)|\leq 1$ and
 $d\rho(\lambda)$ is a positive probability measure, $ \int d\rho(\lambda)=1$.
The parameters $\lambda$ are interpreted as hidden variables in a
realist theory.

One has the following  Bell-Clauser-Horn-Shimony-Holt (CHSH)
inequality
\be
\label{eqn3}
 |P(a,b)-P(a,b')+P(a',b)+P(a',b')|\leq 2
\ee
 From the other hand there are such vectors
$(ab=a'b=a'b'=-ab'=\sqrt{2}/2)$ for which one has
\be
\label{eqn4}
 | E_{spin}(a,b)- E_{spin}(a,b')+ E_{spin}(a',b)+ E_{spin}(a',b')|=2\sqrt{2}
\ee Therefore if one supposes that $ E_{spin}(a,b)=P(a,b)$ then
one gets the contradiction.

\section {Criterium of Locality}

In the previous section the space part of the wave function of the
particles was neglected. However exactly the space part is
relevant to the discussion of locality. The complete wave function
is $\psi =(\psi_{\alpha\beta}({\bf r}_1,{\bf r}_2))$
where $\alpha$ and $\beta $ are spinor indices and ${\bf r}_1$ and
${\bf r}_2$ are vectors in three-dimensional space.

We suppose that detectors are located within the two localized
regions ${\cal O}_1$ and ${\cal O}_2$ respectively, well separated
from one another. Quantum correlation describing the measurements
of spins at the localized detectors is
\be
\label{eqn6}
 E(a,{\cal O}_1,b,{\cal O}_2)=\left<\psi| \sigma\cdot a   P_{{\cal O}_1}
 \otimes  \sigma\cdot b  P_{{\cal O}_2} |\psi\right>
 \ee
Here $P_{{\cal O}}$ is the projection onto the region ${\cal O}$.
Let us consider the case when the wave function has the form
$\psi=\psi_{spin}\phi({\bf r}_1,{\bf r}_2)$. One has
\be
\label{eqn7}
 E(a,{\cal O}_1,b,{\cal O}_2)=g ({\cal O}_1,{\cal O}_2)
  E_{spin}(a,b)
 \ee
 where the function
\be
\label{eqn8}
 g ({\cal O}_1,{\cal O}_2)=\int_{{\cal O}_1x{\cal O}_2}|\phi({\bf
 r}_1,{\bf
 r}_2)|^2 d{\bf r}_1d{\bf r}_2
 \ee
 describes correlation of particles in space.
 Note that one has
\be
\label{eqn8g}
0\leq g ({\cal O}_1,{\cal O}_2)\leq 1
 \ee
 To investigate the property of locality in a realist theory of hidden variables
  we will study whether
 the quantum correlation (\ref{eqn7}) can be represented in the
 form (\ref{eqn2}). More exactly one inquires whether one can write
 the representation
\be
\label{eqn9}
 g({\cal O}_1,{\cal O}_2)E_{spin}(a,b)=\int A(a,{\cal O}_1,\lambda)
 B(b,{\cal O}_2,\lambda) d\rho(\lambda)
 \ee
We have indicated a possible dependence of $A$ on the region
${\cal O}_1$ and $B$ on ${\cal O}_2$. A possible dependence of the
measure $ d\rho(\lambda)$ on the location of detectors ${\cal
O}_1$ and ${\cal O}_2$ deserves a further discussion.

Note that if we set $ g({\cal O}_1,{\cal O}_2)=1$ in (\ref{eqn9})
as it was done in the original proof of Bell's theorem, then it
means we did a special preparation of the states of particles to
be completely localized inside of detectors. It seems  the
 entanglement of the original states can be destroyed in the
process of such a preparation.

Now from (\ref{eqn3}), (\ref{eqn4}) and (\ref{eqn9}) one obtains
the following {\it criterium of locality} in a realist theory of
hidden variables
\be
\label{eqn10}
 g ({\cal O}_1,{\cal O}_2)\leq 1/\sqrt 2
 \ee
 If the inequality (\ref{eqn10}) is valid for  regions  ${\cal
O}_1$ and ${\cal O}_2$ which are well separated from one another
 then there is no violation
of the CHSH inequalities (\ref{eqn3}) and therefore there is no
violation of locality in the corresponding state $\psi$. From the
other side, if for a pair of well separated regions  ${\cal O}_1$
and ${\cal O}_2$ one has
\be
\label{eqn11}
 g ({\cal O}_1,{\cal O}_2) > 1/\sqrt 2
  \ee
then it could be  violation of the realist locality in these
regions for a given state.

\section {Gaussian Wave Functions}

Now let us consider the criterium of locality for Gaussian wave
functions. We will show that with a reasonable accuracy there is
no violation of locality in this case. Let us take the wave
function $\phi$ of the form $\phi=\psi_{1}({\bf r}_1)\psi_{2}({\bf
r}_2)$  where the individual wave functions have the moduli
\be
\label{eqn12}
|\psi_{1}({\bf
 r})|^2 =(\frac{m^2}{2\pi})^{3/2}e^{-m^2{\bf r}^2/2},
|\psi_{2}({\bf
 r})|^2 =(\frac{m^2}{2\pi})^{3/2}e^{-m^2 ({\bf r}- {\bf l})^2/2}
    \ee
We suppose that  the length of the vector ${\bf l}$ is much larger
than $1/m$. We can make measurements of  $P_{{\cal O}_1}$ and
 $P_{{\cal O}_2}$ for any  well separated regions  ${\cal O}_1$
 and ${\cal O}_2$. Let us suppose a rather nonfavorite case for
 the criterium of locality when the wave functions of the particles are almost
 localized inside the regions ${\cal O}_1$
 and ${\cal O}_2$ respectively. In such a case the function $g ({\cal O}_1,{\cal O}_2)$
 can take values near its maxumum.
 We suppose that the region
${\cal O}_1$ is given by $|r_i|<1/m, {\bf r}=(r_1,r_2,r_3) $ and
the region ${\cal O}_2$ is obtained from ${\cal O}_1$ by
translation on ${\bf l}$. Hence $\psi_{1}({\bf r}_1)$ is a
Gaussian function with modules appreciably different from zero
only in ${\cal O}_1$ and similarly $\psi_{2}({\bf r}_2)$ is
localized in the region ${\cal O}_2$. Then we have
\be
\label{eqn13}
 g ({\cal O}_1,{\cal O}_2)=\left(\frac{1}{\sqrt {2\pi}}\int_{-1}^1
e^{-x^2/2}dx\right)^6
  \ee
One can estimate (\ref{eqn13}) as
\be
\label{eqn14}
 g ({\cal O}_1,{\cal O}_2)< \left(\frac{2}{\pi}\right)^3
  \ee
which is smaller than $1/\sqrt 2$. Therefore the locality
criterium (\ref{eqn10}) is satisfied in this case.

\section {Conclusions}

It is shown in this note that if we do not neglect the space part
of the wave function of two particles then the prediction of
quantum mechanics for the correlation with the Gaussian space
parts of the wave functions can be consistent with Bell's
inequalities. One can say that Einstein's local realism is
restored in this case.

We did only a rough estimate and it would be interesting to
investigate whether one can prepare a reasonable wave function for
which the condition of nonlocality (\ref{eqn11}) is satisfied for
a pair of the well separated regions. In principle the function
 $g ({\cal O}_1,{\cal O}_2)$ can approach its maximal value 1 if the
 wave  functions of the particles are very well localized
 within the detector regions ${\cal O}_1$ and ${\cal O}_2$ respectively.
  However, perhaps to establish
 such a localization one has to destroy the original entanglement
 because it was created far away from detectors. If $D_1$ is a
 region which contains the place where the entangled state was
 created and the path to the detector  ${\cal O}_1$ then the
 conditional
 probability of finding the particle 1 in the region  ${\cal O}_1$
 with the projection of spin along  vector $a$ is given by Bayes's
 formula $P(a,{\cal O}_1|D_1)=P(a,{\cal O}_1,D_1)/P(D_1)$.
 It is especially important if one could prepare such strictly
localized within ${\cal O}_1$ and ${\cal O}_2$ entangled states in
real experiments. Then one could say that "local realism" must be
rejected.

 One has to study the
dependence of the wave function not only on the space variables
but also on time. It seems the function $g ({\cal O}_1,{\cal
O}_2)$ and its relativistic analogue should be taken into account
in discussions of experiments performed in the studying of quantum
nonlocality. In particular it gives a  contribution   to the
efficiency of detectors.

\section{Acknowledgments}

I would like to thank A.Khrennikov for numerous fruitful
discussions and  the Department of Mathematics, Statistics and
Computer Sciences, University of Vaxjo, where this work was done,
for the warm hospitality. This work was presented at the
International conference "Foundations of Probability and Physics"
held in Vaxjo University, Sweden, 27 Nov.- 1 Dec. 2000. I am
grateful to L.Accardi, L.E. Ballentine, D.H.E. Gross, W.M. De
Muynck, A.F. Kracklauer, J.- A. Larsson and J. Summhammer for
comments and critical remarks.


\end{document}